\documentclass[conference]{IEEEtran}
\IEEEoverridecommandlockouts
\usepackage{cite}
\usepackage{amsmath,amssymb,amsfonts}
\usepackage{algorithmic}
\usepackage{graphicx}
\usepackage{textcomp}
\usepackage{xcolor}
\usepackage{enumerate}
\usepackage{float}
\usepackage{listings}
\usepackage{environ}
\usepackage{color}
\usepackage{fancyvrb}
\usepackage{csquotes}
\def\BibTeX{{\rm B\kern-.05em{\sc i\kern-.025em b}\kern-.08em
    T\kern-.1667em\lower.7ex\hbox{E}\kern-.125emX}}
\begin{document}

\title{Automated CVE Analysis: Harnessing Machine Learning In Designing Question-Answering Models For Cybersecurity Information Extraction\\
}

\author{\IEEEauthorblockN{Tanjim Bin Faruk}
\IEEEauthorblockA{\textit{Computer Science Department} \\
\textit{Colorado State University}\\
Fort Collins, CO \\
tanjim@colostate.edu}
}

\maketitle

\begin{abstract}
The vast majority of cybersecurity information is unstructured text, including critical data within databases such as CVE, NVD, CWE, CAPEC, and the MITRE ATT\&CK Framework. These databases are invaluable for analyzing attack patterns and understanding attacker behaviors. Creating a knowledge graph by integrating this information could unlock significant insights. However, processing this large amount of data requires advanced deep-learning techniques. A crucial step towards building such a knowledge graph is developing a robust mechanism for automating the extraction of answers to specific questions from the unstructured text. Question Answering (QA) systems play a pivotal role in this process by pinpointing and extracting precise information, facilitating the mapping of relationships between various data points. In the cybersecurity context, QA systems encounter unique challenges due to the need to interpret and answer questions based on a wide array of domain-specific information. To tackle these challenges, it is necessary to develop a cybersecurity-specific dataset and train a machine learning model on it, aimed at enhancing the understanding and retrieval of domain-specific information. This paper presents a novel dataset and describes a machine learning model trained on this dataset for the QA task. It also discusses the model's performance and key findings in a manner that maintains a balance between formality and accessibility.
\end{abstract}

\begin{IEEEkeywords}
Cybersecurity, Question Answering, Dataset, Machine Learning\end{IEEEkeywords}

\section{Introduction}

In recent years, there has been a significant proliferation of information related to cybersecurity, driven by the rapid replacement of older technologies with new innovations and an increasing reliance on these technologies. This shift has substantially expanded the attack surface, leading to a rise in cyber incidents. High-profile attacks such as the WannaCry ransomware, NotPetya, the Equifax data breach, and the Log4j vulnerability represent only a fraction of the emerging threats. Furthermore, the expansion of Internet of Things (IoT) devices, enhanced cloud connectivity, and advancements in AI and ML technologies have further enlarged this attack surface.

Cybersecurity professionals and organizations are intensifying their efforts to collect and analyze data to enhance their understanding of vulnerabilities and streamline the discovery process. Security databases such as the NVD CVE, which catalogs common vulnerabilities and exposures, the MITRE ATT\&CK Framework \cite{strom2018mitre}, which details attacker tactics and techniques, and others like CWE and CAPEC, are invaluable for gaining insights into both current and potential future threats. However, these databases often contain unstructured information, which poses a challenge for automated processing—a crucial capability in cybersecurity, where time-sensitive decisions are necessary as threats unfold.

To address this issue, developing question-answering models trained on cybersecurity database data could be immensely beneficial. These models aim to automate the extraction of critical information and insights, significantly aiding cybersecurity experts by saving time and enhancing the efficiency of their response to cyber threats.

With the proliferation of AI/ML technologies, numerous pre-trained models such as BERT, GPT-2, and T5 have become available. These models, trained on extensive, domain-agnostic datasets like the Wikipedia corpus, are versatile enough to be adapted for various downstream tasks including Named Entity Recognition (NER), Sentiment Analysis, and Masked Language Modeling. However, while these pre-trained models possess broad applicational potential, their effectiveness diminishes in domain-specific scenarios due to their initial training on general data.

In the context of cybersecurity, this limitation is pronounced as these models lack exposure to sector-specific terminology such as \textit{Ransomware}, \textit{OAuth}, \textit{API}, and \textit{Keylogger}. Moreover, the challenge is compounded by homographs—words that, while spelled the same in regular English, carry distinct meanings within the cybersecurity domain, such as \textit{Honeypot}, \textit{Patch}, \textit{Handshake}, and \textit{Worm}. Therefore, there is a critical need to further train these models with domain-specific data to enhance their utility and accuracy in cybersecurity applications.

This study introduces a novel QA CVE dataset specifically designed for the training of pre-trained machine learning models using cybersecurity-focused data. The aim is to develop a question-answering model capable of effectively responding to queries based on descriptions of security vulnerabilities. The key contributions of this study are summarized as follows:

\begin{itemize}
    \item A novel annotated dataset tailored for training machine learning models on data specific to the cybersecurity domain.
    \item The development of a question-answering pipeline that utilizes this dataset to extract information from unstructured security vulnerability descriptions.
\end{itemize}

The remainder of this paper is structured as follows: Section II presents a review of the literature relevant to this
topci. Section III delineates the background information
necessary for understanding the context of the study. Section IV elaborates on the methodology employed. Section
V outlines the experimental setup. Section VI describes the
evaluation metrics utilized in this research. Sections VII and VIII detail the principal findings of the study. Section IX proposes potential directions for future research within this project. Finally, Section X summarizes the study, discussing its current limitations and suggesting avenues for future inquiry.

\section{Related Work}

\cite{cybert} developed a cybersecurity corpus from open-source, unstructured, and semi-structured Cyber Threat Intelligence (CTI) data, which they employed to fine-tune a base BERT model using Masked Language Modeling (MLM) to identify specialized cybersecurity entities. However, the authors did not extend their work to implement downstream tasks such as named entity recognition or question answering using their dataset.

Similar to the approach taken by \cite{cybert}, \cite{securebert} developed a domain-specific language model named SecureBERT, which was trained on cybersecurity threat intelligence reports. This training utilized automated labeling techniques facilitated by semantic role labeling. However, the authors limited the model's availability to Masked Language Modeling capabilities and did not extend its application to question-answering (QA) downstream tasks.

\cite{AEEOCD} conceptualized each cybersecurity vulnerability as an event, representing it through a combination of attributes including the cause, attacker, consequence, operation, location, and version. They introduced a model named VE-Extractor, designed to extract these event-based representations from vulnerability descriptions. Initially, the authors applied Named Entity Recognition (NER) techniques to label the data. However, this approach has inherent limitations, notably its inability to represent nested relationships where label overlaps occur. Additionally, the scope of their work was restricted by the use of a limited set of labels.

\cite{cvejoin} aggregated data from over 200,000 vulnerabilities across a broad range of security databases, providing a valuable resource for various research areas such as risk analysis of security flaw exploitation and vulnerability severity prediction. Despite the impressive volume of data, the dataset has notable limitations, including the absence of detailed vulnerability characteristics and associated threat intelligence.

\cite{ALOTEOAVRCWN} employed Named Entity Recognition (NER) to fine-tune a BERT model using CVE descriptions. The range of labels used by the authors was limited, and notably, the model exhibited particularly poor performance on the \textit{Root Cause} label, which is considered the most critical label in the labeling task.

\section{Background Information}

\subsection{CVE}

The Common Vulnerabilities and Exposures (CVE) system, established in 1999, serves as a public database that catalogs information security vulnerabilities. Each vulnerability listed is assigned a unique CVE identifier, facilitating a standardized method for security professionals, businesses, academics, and other stakeholders to share data concerning cybersecurity issues. The system supports organizations in their vulnerability management efforts by integrating CVE identifiers with the Common Vulnerability Scoring System (CVSS) scores to aid in prioritization and planning.

Vulnerabilities refer to flaws within a system that, if exploited, allow cybercriminals to execute unauthorized actions, such as running malicious code, accessing system memory, or installing malware. These vulnerabilities might enable attackers to alter, delete, or steal sensitive information. On the other hand, Exposure is defined as an error that inadvertently provides attackers access to a system or network, potentially leading to significant consequences like data breaches or the illicit sale of personally identifiable information (PII) on the dark web.

The CVE initiative is coordinated by the National Cybersecurity FFRDC (Federally Funded Research and Development Center), which is managed by the MITRE Corporation and funded by the US government through the Department of Homeland Security (DHS) and the Cybersecurity and Infrastructure Security Agency (CISA). As a publicly accessible resource, CVE is available at no cost, supporting a wide array of cybersecurity activities globally.

\subsection{Pretrained Model, Transfer Learning and Fine Tuning}

Pre-training in neural networks provides a foundational advantage by leveraging previously gained insights to solve new, but related, problems. This approach is widely beneficial, including in areas such as cybersecurity.

For instance, imagine a machine learning model, \enquote{SecNet}, specifically developed to identify phishing emails. Once trained and optimized, SecNet's parameters are stored. When faced with a new challenge, like detecting malicious URLs, instead of building a completely new model, we can apply transfer learning techniques to adapt SecNet to this task. By retraining it with a dataset of URLs, SecNet uses its prior experience with phishing emails to learn and identify patterns associated with malicious web activity.

In essence, a pre-trained model like SecNet serves as a valuable starting point. Originally designed for large-scale tasks such as email filtering, it can be repurposed using transfer learning to extend its capabilities to new cybersecurity challenges, such as URL classification. This method not only saves time and resources but also enhances the model's effectiveness by building on pre-existing knowledge.

\section{Methodology}

\subsection{Dataset}

\subsubsection{Procurement}

The procurement of CVEs from the NVD website was guided by specific criteria to ensure a diverse and relevant dataset:

\begin{itemize}
    \item Published within the last five years, spanning from 2020 to 2024
    \item Encompassed a variety of vendors and software applications
    \item Included different types of vulnerabilities and impacts
    \item Featured a broad spectrum of CVSS scores
    \item Descriptions varied in length
\end{itemize}

Following these parameters, we selected $200$ CVEs from each year between 2020 and 2024, culminating in a dataset of $1000$ CVEs. Out of this collection, $200$ CVEs from the years 2023 and 2024 were specifically utilized for annotation and demonstration purposes. The following information was collected for the shortlisted CVEs:

\begin{itemize}
    \item CVE-ID
    \item Description
    \item Published Date
    \item CVSS Version $3.x$ Score
\end{itemize}

\subsubsection{Labels}

In developing our data model, we established a set of 16 comprehensive labels to ensure the training encompasses both breadth and depth. These labels are designed to capture a wide array of attributes associated with each vulnerability, as described below:

\begin{enumerate}[(i)]
    \item \textbf{Vendor}: This label identifies the company that developed the software, such as Microsoft, Apple, or Google.
    \item \textbf{Software}: This refers to the specific product or service provided by the vendor, which may consist of multiple components, such as MS Word or Google Home.
    \item \textbf{Software Version}: This denotes the specific version of the software where the vulnerability exists, examples include versions \enquote{before 1.0}, \enquote{after 3.1.2}, or \enquote{through 8.7-8.9}.
    \item \textbf{Operating System}: This category includes operating systems on which the software runs, such as Windows, Linux, MacOS, Android, iOS, Unix, ChromeOS, and Ubuntu.
    \item \textbf{Source}: This specifies the exact location, function, or component within the software environment that harbors the vulnerability, such as an Authentication Module or Image Parsing Library.
    \item \textbf{Trigger}: This label describes the event or condition that triggers the vulnerability, such as processing malformed input.
    \item \textbf{Reason}: This identifies the underlying cause of the vulnerability, such as Improper Input Validation, Race Conditions, or Memory Management Errors.
    \item \textbf{System State}: This includes any preconditions necessary for the attack to succeed, like a specific environment variable setting.
    \item \textbf{Consequences}: This details the potential outcomes of the vulnerability, which could include Data Leakage, Identity Theft, Privilege Escalation, or Root Access.
    \item \textbf{Vulnerability Type}: This classifies the type of vulnerability, such as SQL, XSS, or SSRF.
    \item \textbf{Attacker Action}: This describes any action an attacker must take to exploit the vulnerability, like crafting a special packet or manipulating input data.
    \item \textbf{Network Access}: This denotes whether Local or Remote network access is required to exploit the vulnerability.
    \item \textbf{Privilege}: This label categorizes the authentication level required to exploit the vulnerability, with possible values including Unauthenticated, Authenticated, or Simple.
    \item \textbf{User Interaction}: This indicates whether and what type of user interaction is necessary for the exploit to succeed, such as requiring the user to be logged in or to click on something.
    \item \textbf{Exploit}: This notes whether a public exploit is available and may include details such as \enquote{Exploit available with VulDB ID $<$XYZ$>$}.
    \item \textbf{Patch}: This identifies whether a patch is available for the vulnerability, potentially including an identifier like \enquote{Patch available with identifier $<$PQR$>$}.
\end{enumerate}

These labels are utilized to annotate our dataset of CVEs systematically, ensuring detailed and structured vulnerability analysis for our research.

\subsubsection{Annotation}

\begin{figure}[H]
    \centering
    \includegraphics[width=0.5\textwidth]{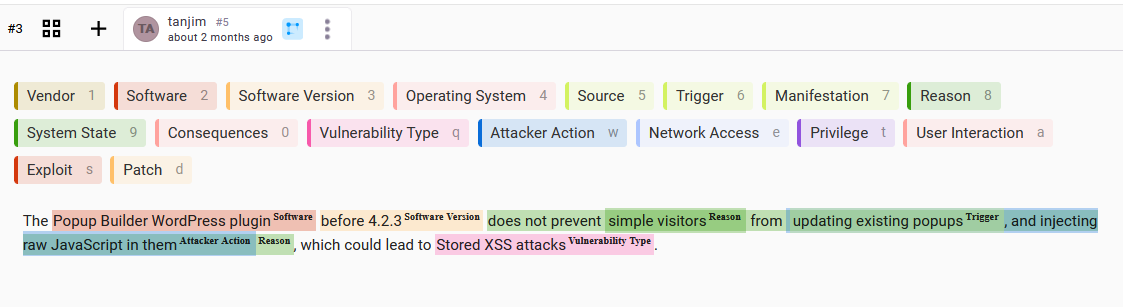}
    \caption{Annotation Example}
    \label{fig:annotation}
\end{figure}

The annotation process was conducted by a team of two expert annotators, who followed the previously outlined guidelines. To facilitate a collaborative approach to annotation, an open-source tool, \cite{label-studio}, was employed. The CVEs were evenly distributed between the annotators, with each responsible for annotating 200 CVEs out of a total of 400 assigned CVEs. An illustration of an example annotation is provided in Figure \ref{fig:annotation}.

\subsubsection{Labels to Question}

The annotation process was initially approached as a Named Entity Recognition (NER) task, where annotators identified specific text regions within the vulnerability descriptions and assigned them predefined labels. However, for the purpose of training machine learning models, these labels were transformed into a question-answering (QA) format. This conversion involved framing each label as a question, providing a context that encapsulates the vulnerability, and defining an explicit answer extracted from the text.

Below is a list of the mappings from labels to corresponding questions utilized in our dataset, facilitated through a code block:

\begin{Verbatim}[frame=lines,framesep=3mm]
label_to_question = {
    "Vendor": "Who is the 
    vendor involved?",
    "Software": "What software 
    is vulnerable?",
    "Software Version": "Which versions 
    of the software are affected?",
    "Operating System": "Which operating 
    system  is mentioned?",
    "Source": "What is the source 
    component of the vulnerability?",
    "Trigger": "What action or condition 
    triggers the vulnerability?",
    "Reason": "Why does the vulnerability 
    exist?",
    "System State": "What system state 
    allows for the vulnerability to be 
    exploited?",
    "Consequences": "What are the 
    potential consequences of the 
    vulnerability?",
    "Vulnerability Type": "How is the 
    vulnerability classified?",
    "Attacker Action": "What must an 
    attacker do to exploit 
    the vulnerability?",
    "Network Access": "What type of 
    network access does an 
    attacker need?",
    "Privilege": "What level of privilege 
    is required for the attack?",
    "User Interaction": "Does the exploit 
    require user interaction?",
    "Exploit": "Is there a public exploit 
    available for the vulnerability?",
    "Patch": "Has a patch been issued 
    for the vulnerability?",
}
\end{Verbatim}

The transformation from NER to QA format addresses the challenge of nested entities, where a single text region may overlap across multiple labels, which NER cannot accommodate. The QA approach allows for a more nuanced representation of data, enabling more complex interactions to be modeled effectively.

For illustrative purposes, an example of how the annotated data appears in the QA format is provided below:

So, for the annotation example shown in \ref{fig:annotation}, the following would be how the training data look like:

\begin{Verbatim}[frame=lines,framesep=3mm]
{
    'context': 'The Popup Builder 
    WordPress plugin before 4.2.3 
    does not prevent simple visitors 
    from updating existing popups, 
    and injecting raw JavaScript 
    in them, which could lead to 
    Stored XSS attacks.', 

    'question': 'Which versions of the 
    software are affected?', 
    'answer': {
            'text': 'before 4.2.3', 
            'start': 35, 
            'end': 47
    }
}

....
\end{Verbatim}

This format not only provides a clear delineation of data points but also facilitates a more effective training of machine learning models for question-answering tasks within the cybersecurity domain.

\subsection{Model Selection Criteria}

For the purpose of selecting suitable models to train on our question-answering (QA) dataset, we established a set of criteria to ensure optimal compatibility and performance:

\begin{itemize}
    \item The model must be trained on a diverse dataset to enhance its adaptability to varied inputs.
    \item The model should be specifically suitable for QA tasks, indicating a proven capability in understanding and processing natural language queries.
    \item Ideally, the model should have prior training on a QA-specific dataset, providing foundational knowledge relevant to our task.
    \item The model must be compatible with our existing GPU setup, ensuring efficient utilization of computational resources.
    \item There should be a balance between accuracy and resource consumption, making the model practical for sustained use.
\end{itemize}

Based on these criteria, we shortlisted the following models:

\begin{itemize}
    \item \textbf{BERT (Bidirectional Encoder Representations from Transformers)}: A pioneering model in NLP, BERT’s architecture allows it to understand the context of a word based on all of its surroundings (left and right of the word). The model’s deep bidirectional nature makes it highly effective for QA tasks where context is crucial. BERT was included due to its robust performance across various NLP benchmarks, including QA.\cite{devlin-etal-2019-bert}
    \item \textbf{DistilBERT}: This model is a distilled version of BERT that retains most of the original model’s effectiveness but with fewer parameters and faster training times. DistilBERT’s efficiency in handling large datasets with reduced computational resources made it a suitable candidate for our GPU constraints.\cite{distilbert}
    \item \textbf{RoBERTa (Robustly Optimized BERT approach)}: An optimized version of BERT, RoBERTa modifies key hyperparameters in BERT and removes the next-sentence pretraining objective. Its enhancements lead to improved performance on NLP tasks, which is advantageous for complex QA scenarios in cybersecurity.\cite{roberta}
    \item \textbf{XLNet}: Utilizing a permutation-based training approach, XLNet outperforms BERT on several NLP benchmarks by learning bidirectional contexts better. This model falls under the autoregressive category, where each token can attend to all positions within a permutation. Its inclusion was due to its advanced capabilities in understanding context, crucial for interpreting cybersecurity texts.\cite{xlnet}
    \item \textbf{T5 (Text-to-Text Transfer Transformer)}: T5 converts all NLP problems into a unified text-to-text format, which simplifies the training process and enhances model versatility. Its sequence-to-sequence framework is well-suited for generating answers to questions based on the provided context, a key requirement for QA in cybersecurity.\cite{t5}
\end{itemize}

This selection covers a broad spectrum of model architectures, ensuring that we can evaluate and compare different types of models to determine which best suits our QA needs in the cybersecurity domain.

\section{Experimental Setup}

This section outlines the methodology employed for training the models on our dataset, using the Huggingface Transformers \cite{wolf-etal-2020-transformers} and PyTorch \cite{pytorch} libraries. The experiments were conducted on a machine equipped with an NVIDIA GeForce RTX 3070Ti GPU.

\subsection{Data Preparation and Training}

The dataset was partitioned into an 80-20 split, with 80\% used for training and the remaining 20\% for validation purposes. Tokenization of the text was performed in batches to enhance processing speed, a crucial adjustment given the lengthy nature of vulnerability descriptions. To accommodate the models with maximum length constraints, such as BERT, the maximum token length was set to 384. To avoid loss of critical information due to truncation, overlapping chunks were created using a stride of 128.

During tokenization, the start and end positions of the answers were identified for segments containing the entire answer; if the answer was not fully contained, these positions were set to zero. This approach ensures that the models are trained accurately on the relevant text segments.

\subsection{Model Training Configuration}

The models were trained using the following hyperparameters:
\begin{itemize}
    \item \textbf{Learning Rate}: $2 \times 10^{-5}$
    \item \textbf{Epochs}: $3$
    \item \textbf{Weight Decay}: $0.01$
\end{itemize}

To increase the efficiency of the training process, mixed precision training was enabled. This technique reduces the computational overhead, allowing for faster training times without compromising the model's performance.

\subsection{Validation and Metrics}

For performance evaluation, we used the \textit{evaluate} library to employ the \textit{squad} metric, which is based on the SQuAD \cite{rajpurkar-etal-2016-squad} benchmark. This metric is pivotal for assessing the effectiveness of QA models, focusing on exact matches and F1 scores rather than loss, which provides a more relevant measure of performance in QA tasks.

During validation, the model's output logits were used to determine the predicted start and end positions of the answers. These were then compared with the ground truth positions to calculate the exact match and F1 scores, thus providing a quantitative measure of the model's accuracy and ability to generalize to unseen data.

\section{Evaluation Metrics}

In the domain of models trained on CVE (Common Vulnerabilities and Exposures) descriptions for a question-answering task, selecting the appropriate evaluation metrics is paramount for accurately assessing their performance. These models are tasked with extracting specific text spans from CVE descriptions that answer posed questions, necessitating metrics that measure the accuracy of these extractions effectively. Therefore, we have opted for Exact Match (EM) and F1 Score as our principal evaluation metrics, as they are well-suited to gauge both the precision and the contextual accuracy of the answers provided by the model.

\subsection{Exact Match (EM)}

Exact Match is a strict metric that evaluates whether the predicted answer is identical to the correct answer listed in the dataset. This metric does not allow for any variation in the answer text; any deviation from the annotated answer results in a score of zero, regardless of how close the predicted answer might be to conveying the same meaning. For example, if the question posed is \enquote{What protocol vulnerability does CVE-2022-1234 relate to?} and the model's response is \enquote{HTTP protocol}, it would receive an EM score of $0$ if the documented correct answer is \enquote{Hypertext Transfer Protocol (HTTP)}.

EM is highly relevant to our task because it quantifies the model’s capability to accurately identify and return the exact text span corresponding to the answer within the CVE descriptions. This metric is indispensable for understanding how often the model exactly matches the defined answers in our dataset.

\subsection{F1 Score}

The F1 Score, more forgiving than EM, assesses the overlap of words between the predicted and the correct answer, providing a measure of partial accuracy. This metric computes the harmonic mean of precision (the proportion of words in the predicted answer that are found in the correct answer) and recall (the proportion of words in the correct answer that are also in the predicted answer), thereby offering a balanced view of the model's performance.

The F1 Score is particularly significant for our evaluations as it accounts for instances where the model captures the essence of the answer but does not align exactly with the predefined answer boundaries. It appreciates semantic closeness in responses, which is crucial when answers may contain synonyms, rephrased wording, or additional contextually pertinent words. For example:

\begin{Verbatim}[frame=lines]

Question: What type of vulnerability 
is described in CVE-2022-3052?

Documented Correct Answer: 
Cross-site scripting (XSS)

Model's Predicted Answer: 
XSS vulnerability
    
\end{Verbatim}

In this example, an exact match would give a $0$ score, but the F1 score will be $0.5$.

By incorporating both Exact Match and F1 Score into our evaluation strategy, we can thoroughly assess the performance of our model on the CVE description-based question-answering task. EM provides an absolute measure of precision in identifying specific answer spans, while the F1 Score gives insights into the semantic and contextual accuracy of the answers. This dual approach ensures that our evaluation captures both the precision in text span retrieval and the quality of content comprehension by the model, which are essential for applications in cybersecurity.

\section{Results}

The dataset for the experiment consisted of 304 question-context pairs. Following an 80-20 split, 243 pairs were allocated for training, with the remaining pairs used for validation.

\begin{table}[h]
\setlength{\abovecaptionskip}{-10pt} 
\setlength{\belowcaptionskip}{10pt} 
\resizebox{\columnwidth}{!}{
\scriptsize 
\begin{tabular}{|c|c|c|}
\hline
\textbf{Model}        & \textbf{Exact Match} & \textbf{F1}\\ \hline
bert-base-uncased  & 1.64  & 12.35  \\ \hline
bert-base-cased    & 1.64  & 12.71   \\ \hline
xlnet-base-cased  &  1.64 &  13.15 \\ \hline
distilbert-base-cased-distilled-squad & 39.34 & 65.42 \\ \hline
deepset/roberta-base-squad2 & 65.57 & 80.24 \\ \hline
\end{tabular}
}
\newline
\end{table}

After multiple experimental runs, the highest F1 score achieved was 80.24, while the exact match score reached 65.57 using the \enquote{deepset/roberta-base-squad2} model.

\section{Result Analysis}

Despite the constraints posed by a limited training dataset, models that were fine-tuned on a question-answering dataset akin to SQuAD demonstrated superior performance compared to those that were not. This observation underscores the importance of selecting models that have been specifically fine-tuned on similar datasets for achieving optimal results.

The performance of the non fine-tuned models on the validation dataset was unsatisfactory, attributable to several identifiable factors:

\begin{itemize}
    \item \textbf{Insufficient Annotated Data}: The primary limitation was the size of the training dataset, which comprised only 243 question-context pairs. This quantity proved inadequate for training the model effectively on the complex patterns of relationships between questions and contexts derived from CVE descriptions. A larger dataset with more annotated examples would likely enhance the model's ability to learn these patterns more comprehensively, thereby improving its performance metrics, such as the exact match and F1 scores, on the validation dataset.
    
    \item \textbf{Custom Tokenizer}: The performance issues may also stem from inadequate tokenization due to the vocabulary limitations of the pre-trained base model. The base model's vocabulary lacked sufficient cybersecurity-specific terminologies, potentially leading to ineffective tokenization. This deficiency could have negatively impacted the model's training and subsequent prediction accuracy. Developing a tokenizer tailored to the cybersecurity domain, using the training dataset, might allow for more accurate modeling of the data and improved prediction outcomes.
\end{itemize}

\section{Future Work}

To enhance the model's performance, future efforts will focus on expanding the annotated dataset. Additionally, to accommodate the increased dataset size and accelerate the training process, distributed training or multi-GPU setups will be implemented. Moreover, to better capture the unique terminology of the cybersecurity domain, a custom tokenizer will be developed and trained on the annotated dataset. These strategies are intended to refine the model's understanding and handling of cybersecurity-specific language, thereby improving its predictive accuracy and efficiency.

\section{Conclusion}

This study set out to enhance the automated processing of cybersecurity information by leveraging machine learning models trained on a novel question-answering (QA) dataset derived from CVE descriptions. Our project has demonstrated both the potential and the limitations inherent in applying advanced NLP techniques to cybersecurity data, a domain rich with complex and specialized information.

Our findings indicate that while pre-trained models like BERT and its variants hold promise for this type of application, the unique challenges posed by the cybersecurity domain—such as the need for domain-specific vocabulary and a deeper contextual understanding—require more tailored approaches. The performance of our models on the CVE dataset, particularly measured through metrics such as the Exact Match and F1 Score, underscored the necessity for extensive domain-specific training data. The low scores observed in our experiments highlight the need for a larger and more diverse dataset to effectively capture the nuances of cybersecurity texts.

Moreover, the experience gathered from the annotation process and subsequent model training and validation phases suggests several paths forward. Developing a custom tokenizer that can better handle the specialized terminology found in CVE entries is likely to yield significant improvements in model performance. Additionally, employing techniques such as transfer learning and possibly training models from scratch with a focus on cybersecurity texts could further enhance their effectiveness.

Looking ahead, this work paves the way for a series of incremental improvements in the automation of cybersecurity information processing. Expanding the dataset, refining the training process, and continuously updating the models with new data as it becomes available are critical steps. Furthermore, future research could explore the integration of these QA models into broader cybersecurity systems, such as automated threat detection platforms or real-time security incident response frameworks, thereby contributing to more robust and intelligent cybersecurity defenses.

In conclusion, the project has provided valuable insights into the capabilities of current machine learning technologies when applied to the cybersecurity domain and has laid the groundwork for future advancements that could significantly impact how cybersecurity data is processed and utilized.

\bibliographystyle{IEEEtran}
\bibliography{bibliography}

\end{document}